\documentclass[prb,reprint,amsfonts,superscriptaddress]{revtex4-1}

\pdfoutput=1 
\usepackage{amsmath}
\usepackage{amssymb}
\usepackage{amsfonts}
\usepackage[pdftex]{graphicx}
\usepackage{units}
\usepackage[usenames]{color}
\usepackage{dcolumn}
\usepackage{bm}
\usepackage{float}
\usepackage{xspace}
\usepackage[usenames]{color}
\usepackage{textcomp}
\usepackage{multirow}
\usepackage[]{placeins}

\def\Ef{$E_{\rm F}$}
\def\Eb{$E_{\rm B}$}

\def\kpara{{\bf k}$_\parallel$}
\def\kperp{{\bf k}$_\perp$}

\def\kperp{{\bf k}$_\perp$}
\def\invA{\AA$^{-1}$}

\def\GbarKbar{$\overline{\Gamma}$-$\overline{\rm K}$}

\def\aSn{$\alpha$-Sn}

\newcommand{\celcius}{$ \,^{\circ}\mathrm{C} $\xspace}     
\newcommand{\angs}{\textup{\AA}\xspace}   
\newcommand{\GSM}{$ \mathrm{\Gamma_{7}^{-}} $\xspace} 
\newcommand{\GEP}{$ \mathrm{\Gamma_{8}^{+}} $\xspace} 

\begin{document}
\title{Tailoring the topological surface state in ultrathin \aSn\ (111) films}

\author{V. A. Rogalev}
\affiliation{\mbox{Physikalisches Institut and W\"urzburg-Dresden Cluster of Excellence  ct.qmat, Universit\"at W\"urzburg, 97074 W\"urzburg, Germany}}

\author{F. Reis}
\affiliation{\mbox{Physikalisches Institut and W\"urzburg-Dresden Cluster of Excellence  ct.qmat, Universit\"at W\"urzburg, 97074 W\"urzburg, Germany}}

\author{F. Adler}
\affiliation{\mbox{Physikalisches Institut and W\"urzburg-Dresden Cluster of Excellence  ct.qmat, Universit\"at W\"urzburg, 97074 W\"urzburg, Germany}}

\author{M. Bauernfeind}
\affiliation{\mbox{Physikalisches Institut and W\"urzburg-Dresden Cluster of Excellence  ct.qmat, Universit\"at W\"urzburg, 97074 W\"urzburg, Germany}}

\author{J. Erhardt}
\affiliation{\mbox{Physikalisches Institut and W\"urzburg-Dresden Cluster of Excellence  ct.qmat, Universit\"at W\"urzburg, 97074 W\"urzburg, Germany}}

\author{A. Kowalewski}
\affiliation{\mbox{Physikalisches Institut and W\"urzburg-Dresden Cluster of Excellence  ct.qmat, Universit\"at W\"urzburg, 97074 W\"urzburg, Germany}}

\author{M. R. Scholz}
\affiliation{\mbox{Physikalisches Institut and W\"urzburg-Dresden Cluster of Excellence  ct.qmat, Universit\"at W\"urzburg, 97074 W\"urzburg, Germany}}

\author{L. Dudy}
\affiliation{\mbox{Physikalisches Institut and W\"urzburg-Dresden Cluster of Excellence  ct.qmat, Universit\"at W\"urzburg, 97074 W\"urzburg, Germany}}

\author{L. B. Duffy}
\affiliation{Clarendon Laboratory, Physics Department, Oxford University, OX1~3PU, United Kingdom}

\author{T. Hesjedal}
\affiliation{Clarendon Laboratory, Physics Department, Oxford University, OX1~3PU, United Kingdom}

\author{M. Hoesch}
\affiliation{Diamond Light Source, Didcot, OX11~0DE, United Kingdom}
\affiliation{DESY Photon Science, Notkestraße 85, D-22607 Hamburg}

\author{G. Bihlmayer}
\affiliation{Peter Gr\"unberg Institut, Forschungszentrum J\"ulich and JARA, 52425 J\"ulich, Germany}

\author{J. Sch{\"a}fer}
\affiliation{\mbox{Physikalisches Institut and W\"urzburg-Dresden Cluster of Excellence  ct.qmat, Universit\"at W\"urzburg, 97074 W\"urzburg, Germany}}

\author{R. Claessen}
\affiliation{\mbox{Physikalisches Institut and W\"urzburg-Dresden Cluster of Excellence  ct.qmat, Universit\"at W\"urzburg, 97074 W\"urzburg, Germany}}


\begin{abstract}
We report on the electronic structure of \aSn\ films in the very low thickness regime grown on InSb(111)A. High-resolution low photon energies angle-resolved photoemission (ARPES) allows for the direct observation of the linearly dispersing 2D topological surface states (TSSs) that exist between the second valence band and the conduction band. The Dirac point of this TSS was found to be 200\,meV below the Fermi level in $10$-nm-thick films, which enables the observation of the hybridization gap opening at the Dirac point of the TSS for thinner films. The crossover to a quasi-2D electronic structure is accompanied by a full gap opening at the Brillouin zone center, in agreement with our density functional theory calculations. We further identify the thickness regime of \aSn\ films where the hybridization gap in TSS coexists with the topologically non-trivial electronic structure and one can expect the presence of a 1D helical edge states.

\end{abstract}
\maketitle

\section{\label{Intro} Introduction}

The low temperature $\alpha$-phase of Sn belongs to a family of materials with a topologically non-trivial electronic band structure \cite{Fu:2007ei}. Due to its mono-elemental nature, and the resulting favorable defect chemistry, \aSn\ has recently attracted considerable interest \cite{Barfuss:2013by,Ohtsubo:2013je,kufner2014topological,RojasSanchez:2016gr,Rogalev:2016tz,Huang2017,barbedienne2018angular,liao2018superconductivity}. In particular, \aSn\ thin films grown by molecular beam epitaxy (MBE) exhibit outstanding quality \cite{barbedienne2018angular,RojasSanchez:2016gr,vail2019growth}. In contrast to $ \mathrm{Bi_{2}X_{3}} $ compounds (X = Se or Te), band inversion in bulk \aSn\ involves the \textit{second} valence band (VB) \GSM and conduction band (CB) \GEP, which reveal \textit{s}-like and \textit{p}-like character, respectively \cite{Zhu2012}. Such band order is also typical of HgTe \cite{berchenko1976mercury, volkov1985two} and some half-Heusler compounds \cite{Chadov2010} and results in fact in a double band inversion with a pair of TSSs of different wavefunction localization character \cite{Rogalev:2016tz}. Most of the substrates available for epitaxial growth provide an in-plane compressive strain for \aSn\ films, which drives them into a Dirac semimetal phase \cite{Rogalev:2016tz,Xu2017} with both TSSs being fully degenerate with bulk states. Yet, previous studies on in-plane compressively strained \aSn\ thin films revealed that the hybridization between the upper TSS, in focus of the present work, and the bulk states is weak due to differences in the orbital composition \cite{Rogalev:2016tz,Scholz2017}. This allows for the observation of a sharp ${E(\vec{k})}$ dispersion of this TSS in spin- and angle-resolved photoemission.

The (111) surface of \aSn\ is of particular interest due to its close relationship to the family of 2D honeycomb lattices in group IV and V high-$Z$ materials, e.g., stanene \cite{molle2017buckled} or bismuthene \cite{Reis2016}. These have been widely investigated, both theoretically and experimentally, as a new platform for utilizing helical spin-polarized topological edge states \cite{li2018epitaxial_review}. Despite numerous reports on the fabrication of stanene on a variety of substrates \cite{zhu2015epitaxial,gou2017strain,Xu2018,Zang2018stanenePbTe,Deng2018:staneneCu111,Yuhara2018:staneneAg111,zheng2019epitaxial}, the experimental studies of the 3D to 2D crossover of the TSS in \aSn\ films have remained scarce. With the reduced thickness, the 3D bulk band structure changes developing gap(s) due to increasing confinement in a quantum well, and the surface- and interface-TSSs (at the \aSn/substrate interface) wavefunctions start to overlap. 
Similar to HgTe/CdTe quantum wells \cite{Bernevig2006,Koenig:2007hs}, the band inversion in \aSn\ can be lost at some certain critical thickness \cite{Ohtsubo:2013je,chou2014hydrogenated,DeCoster2018,li2019quantum} or even show an oscillatory behavior \cite{liu2010oscillatory} depending on thickness. Additionally, if the band inversion remains, the hybridization between surface- and interface-TSSs in thin \aSn\ films could open a gap at the Dirac point (DP). At the same time a strong confinement of the 2D TSS on the side-planes leads to the appearance of the 1D helical edge states. In such case the system often can be classified as a 2D quantum spin Hall insulator (QSHI).

In this paper we report on the electronic structure of ultrathin \aSn\ films epitaxially grown on InSb(111)A, and, in particular, on the evolution of the TSS as a function of thickness exploiting high-resolution ARPES and density functional theory (DFT) calculations which notably include the substrate. We find that, in contrast to \aSn\ on InSb(001), the DP of the TSS for the (111) surface orientation is situated significantly below the Fermi energy for $10$\,nm-thick \aSn\ films. The latter enables the direct observation of the hybridization gap opening at the DP for thinner films. Thus, for a 3-nm-thick film we find a gap in the TSS of the order of $\Delta E^{g}_{peak}$ $\approx$ 200\,meV (peak-to-peak). The electronic structure of the quasi-2D 1-nm-thick films exhibits a full gap opening at the $\Gamma$-point ($\Delta E^{g}_{peak}$ $\gtrsim$ 400\,meV). Our DFT calculations for thin \aSn\ films on InSb(111)A show good agreement with the experimental data and provide evidence for the spin-polarized character of the TSS. Furthermore, we establish that coexistence of the TSS hybridization gap and the topologically non-trivial bandstructure appears in a narrow \aSn\ films thickness range of  $\sim$2 - 10\,nm, at which one can expect also the presence of a 1D helical edge states. In addition, we report on a new (8$ \times $8) surface reconstruction observed in low-energy electron diffraction (LEED) and scanning tunneling microscopy (STM).

\section{\label{Methods} Methods}
\aSn\ thin films were grown by MBE on $n$-doped In-terminated InSb(111) substrates. The 8-effusion cell MBE system is directly attached to the high-resolution ARPES system at beamline I05 at the Diamond Light Source (Didcot, UK), allowing for in-vacuum transfers \cite{Baker:2015}.
The substrates were cleaned by several cycles of Ar ions sputtering and annealing until a clear (2$\times$2)-reconstruction was observed by LEED [see Fig.~\ref{STM}(a)]. During thin film growth, the InSb(111) substrates were held at ambient temperature and the \aSn\ film quality was monitored by reflection high-energy electron diffraction (RHEED). The thickness of the Sn layers was varied by changing the deposition time while keeping the flux from the effusion cell constant.
As in the case of (001)-oriented films \cite{Scholz2017}, the XPS data indicate the presence of In atoms on the (sub)surface of \aSn\ films, which could be a result of In interdiffusion and/or In surface segregation that appears during substrate cleaning.

ARPES measurements have been carried out primarily with $p$ linearly polarized light unless stated otherwise, at varying photon energies at beamline I05. ARPES data measured with $s$-polarized light are shown in the Supplementary \cite{supplement}. The endstation is equipped with a Scienta R4000 hemispherical electron analyzer that provides an ultimate energy and angular resolution of $\sim$5\,meV and 0.1$^{\circ}$, respectively.

STM experiments were performed with an Omicron LT-STM at a base pressure $p<$ 5$\times 10^{-11}$\,mbar ($T$=4\,K) using tungsten tips tested on a Ag(111) single crystal for sharpness and spectroscopic properties.

For the DFT calculations we used the full potential linearized augmented plane wave (FLAPW) method \cite{wimmer1981full} as implemented in the FLEUR code in the thin film geometry \cite{krakauer1979linearized}. In this way electrostatic interactions that occur for polar films in repeated-slab calculations are avoided. The muffin-tin radii for Sn, Sb and In were chosen to be 2.3 a.u., while for H we used 0.9 a.u. The $ 4d $ orbitals of Sn and Sb were included as local orbitals. The plane-wave cutoff was 3.8 a.u.$ ^{-1} $ for the wave-functions and 16.3 a.u.$ ^{-1} $ for the potential. We used a 9 x 9 Monkhorst-Pack k-point grid to sample the Brillouin zone and employed the local density.
We employed the local density approximation to the exchange-correlation potential \cite{vosko1980accurate}. Our models for the 0.7\,nm and 2.5\,nm Sn films included four InSb(111)A substrate layers that were charge-compensated on both sides by +/- 0.25 electronic charges in the virtual crystal approximation to simulate a flat profile of the band edges.
The potential profile was checked by monitoring the layer-dependence of the $ 2s $ core levels.
The 7.1\,nm film was calculated without substrate and in all cases hydrogen was deposited on the upper and lower surface of the film to saturate dangling bonds. 
By relaxation we obtained Sn-H and Sb-H distances of 1.71\,\angs and 1.75\,\angs, respectively, which is a bit longer than in SnH$_{4}$ (1.69\,\angs) and SbH$_{3}$ (1.70\,\angs). Although in the experiment other atomic species might saturate the dangling bonds at the surface, with H-termination we simulate successfully the observed absence of the dangling-bonds states at the Fermi level. 
To achieve a correct band-ordering we applied the DFT+U scheme as described in Ref. \cite{Barfuss:2013by}. The same correction was applied to InSb, where the band gap is also underestimated in DFT. The films were relaxed and spin-orbit coupling was applied in a self-consistent manner \cite{li1990magnetic}.

\section{\label{Surface} Experimental results}

\begin{figure}  [!ht]
	\includegraphics[width=1.0\linewidth]{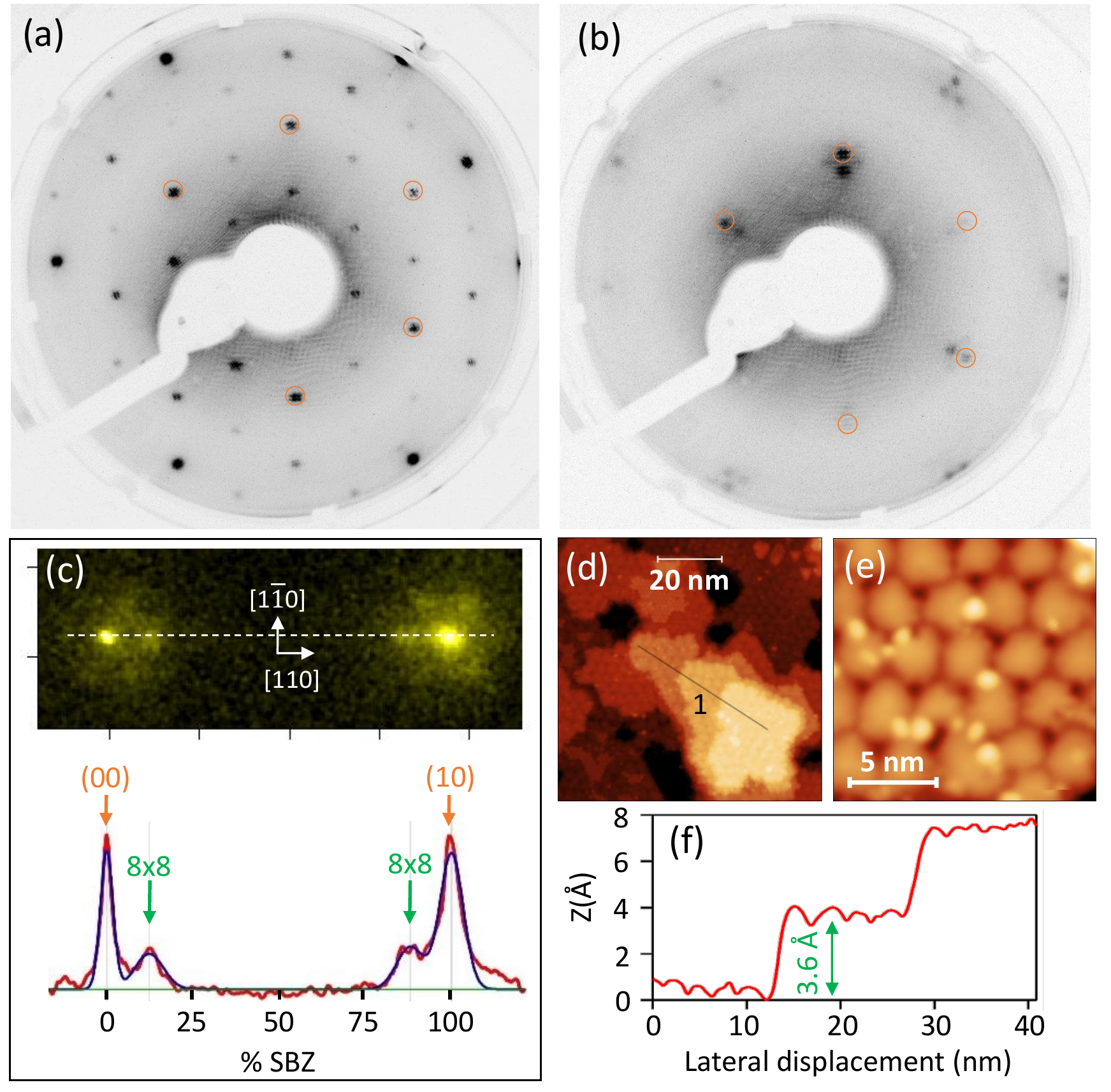}
	\caption{
		(a) LEED taken at $E$=45\,eV on a clean InSb(111)A substrate exhibits (2$\times$2) surface reconstruction (orange circles denote (1$\times$1) spots).
		(b) LEED taken at $E$=45\,eV on  $10$\,nm-thick \aSn(111) film. Apart from the expected (1$\times$1) structure (orange circles), additional spots of high-order surface reconstruction are observed.
		(c) SPA-LEED data taken at $E$=87\,eV and corresponding line profile (data plotted in red, 4-peaks fit - in blue) through (00) and (10) spots allow to assign the new surface reconstruction to be (8$ \times $8).
		(d) STM data measured with $U$=2\,V, $I$=50\,pA at $T$=4\,K.
		(e) STM data on smaller scale showing domains.
		(f) The height profile along path labeled '1' in (d) reveals a step height of 3.6\,\angs.}
	\label{STM}
\end{figure}

During growth, RHEED reveals a well-resolved streak pattern consistent with the substrate symmetry,thus confirming the epitaxial growth of \aSn\ films on InSb(111)A substrate. However, in LEED, apart from the main (1$\times$1) pattern, as-grown \aSn\ films show additional spots stemming from a surface reconstruction: instead of the commonly observed (3$\times$3) surface reconstruction, we find a higher-order spots [Fig.~\ref{STM}(b)]. Spot-profile analysis LEED (SPA-LEED) allowed to identify a (8$\times$8) surface reconstruction [Fig.~\ref{STM}(c)]. It persists when mildly annealing the sample at $T$ $\approx$ 150\,\celcius, leading to sharper LEED and ARPES signals. When the annealing temperature is increased further, the LEED pattern changes to the (1$\times$1) reconstruction, in agreement with other studies \cite{osaka1994surface,fantini2000alpha}. In addition, we performed STM measurements of $10$\,nm-thick \aSn\ films, which are summarized in Figs.~\ref{STM}(d-f). The STM measurements reveal domains with a shape close to hexagonal, which could result from twinning, i.e, an overlap of two 60$^{\circ}$-rotated triangular domains. Twinning was indeed reported for \aSn\ films on Hg$_{0.8}$Cd$_{0.2}$Te(111) substrates \cite{Zimmermann1996Interface}.  The lateral size of domains ranges from 7$\times$ to 8$\times$ lattice constant of unstrained \aSn\ ($a$ = 4.59\,\angs). The step height $h$ $\approx$ 3.6\,\angs is consistent with the interplanar distance between bilayers of {\aSn}(111) films of 3.75\,\angs. To the best of our knowledge, the (8$ \times $8) surface reconstruction was not reported for \aSn\ films so far. Additional STM/STS measurements are necessary to establish the exact structural model of this reconstruction.

\begin{figure}  [!ht]
	\includegraphics[width=\linewidth]{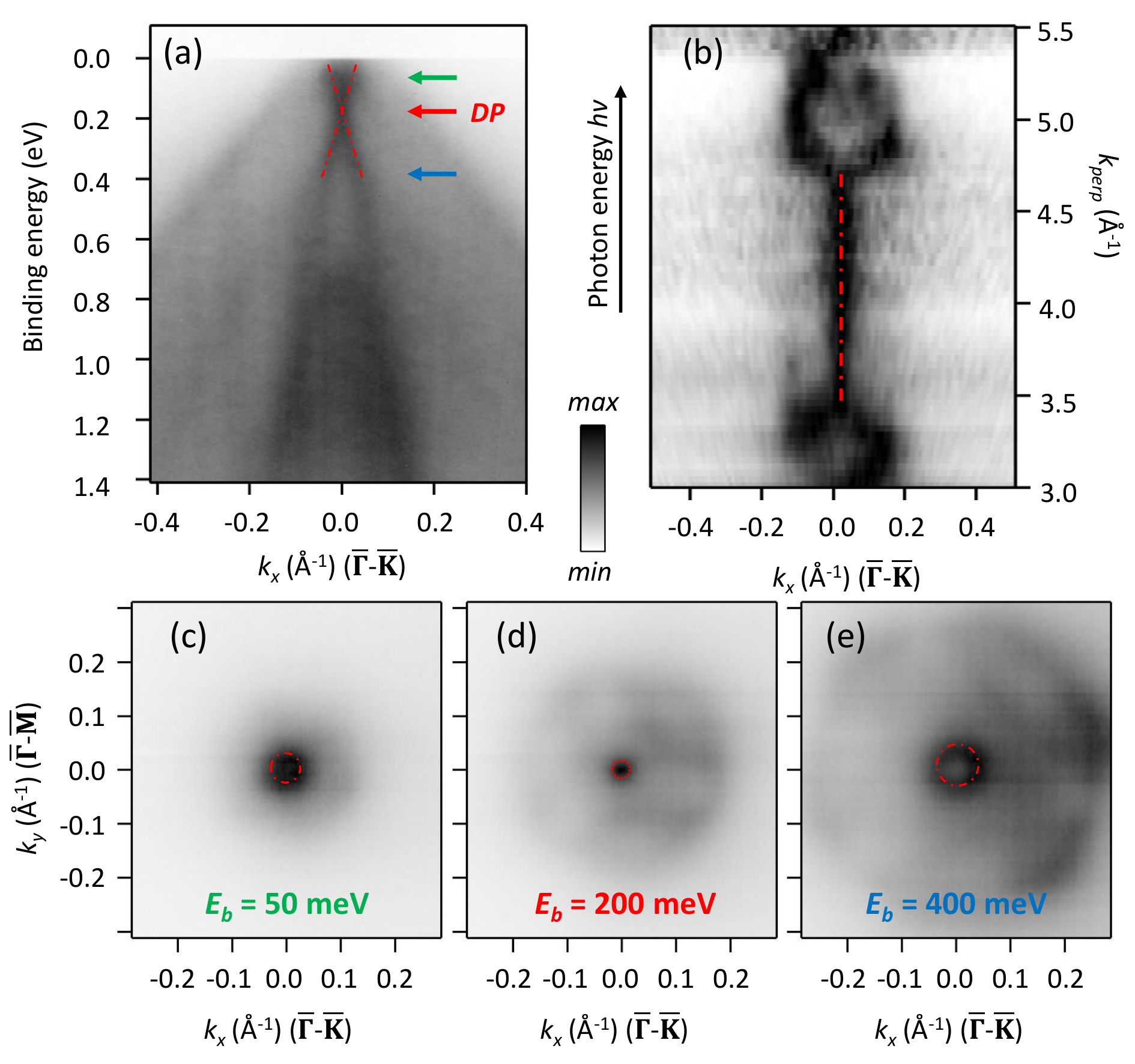}
	\caption{ Experimental electronic structure of a $10$\,nm-thick \aSn\ film on InSb(111)A.
		(a) Band map measured at a photon energy of $h \nu$ = 18\,eV and at $T$ = 8\,K along the \GbarKbar\ direction. (b) Photon energy scan between $h \nu$ = 33\,eV and $h \nu$ = 120\,eV (\kperp $ \approx $ 3\,\invA --  5.5\,\invA), taken with entrance slit oriented along the \GbarKbar\ direction (\Eb=100\,meV). 
		The spectra have been normalized to have equal intensity for each \kperp.
		(c-e) Stacks of experimental constant energy contours at different \Eb. Red dotted lines are guides for the eye indicating the TSS.}
	\label{ARPES_standard}
\end{figure}

\begin{figure}  [!htb]
	\includegraphics[width=\linewidth]{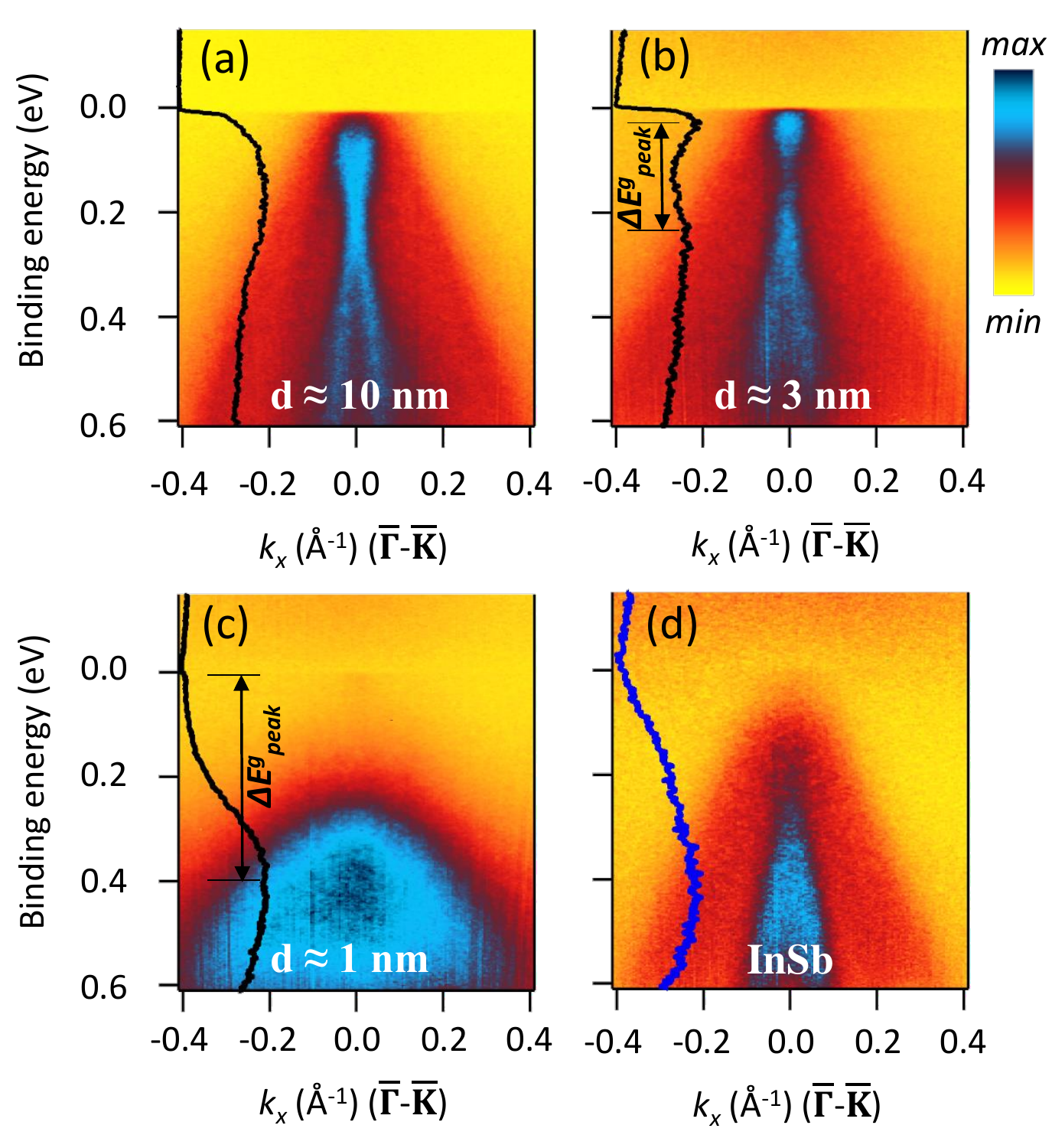}
	\caption{ARPES band maps as a function of \aSn\ film thickness. The photon energy was 18\,eV and the temperature 8\,K. EDCs at normal emission are overlayed on the left of each panel.
		(a) $10$\,nm-thick, (b) $3$\,nm-thick, and, (c) $1$\,nm-thick \aSn\ films. (d) Clean InSb(111)A substrate.}
	\label{Thickness_ARPES}
\end{figure}

Figure \ref{ARPES_standard}(a) shows ARPES data obtained on a $10$\,nm-thick \aSn\ film using a photon energy of $h \nu =$  18\,eV. Such photon energy corresponds to a surface perpendicular momentum \kperp\ = 1.33$\times$(2$ \pi / c $), assuming an inner potential of V$ _{0} =$ 5.8\,eV \cite{Rogalev:2016tz} and $c = $ 3.75\,\angs (interplanar distance between bilayers in \aSn), which allows to highlight the surface states \cite{Rogalev:2016tz,Scholz2017}.
The very presence of a well-ordered surface provides good quality ARPES data, while any signature of the (8$\times$8) reconstruction in the electronic structure is not observed. The Fermi level is pinned $\sim$100\,meV below the valence band maximum of the projected bulk bands [Fig.~\ref{ARPES_standard}(a)], therefore, the expected band gap in the bulk electronic structure defined by quantum confinement \cite{DeCoster2018} is not accessible. Apart from the projected bulk bands that possess 60$^{\circ}$-twinned 3-fold character visible in Fig.~\ref{ARPES_standard}(c-e), the electronic structure of \aSn\ films harbor an additional pair of linear-like crossing bands with a cross-point $\sim$200\,meV below \Ef\ [Fig.~\ref{ARPES_standard}(a)]. The linear-like band has a group velocity $v_{TSS} $=(6.4 $\pm$ 0.5)\,eV\,{\AA}, i.e., (9.7 $\pm$ 0.8) $\times$ 10$^5$\,m/s [see Fig.~\ref{Thickness_ARPES}(a)], which is slightly bigger than previously reported value for \aSn\ \cite{Ohtsubo:2013je,Scholz2017,kufner2014topological}. However, in contrast to \aSn\ on InSb(001) and InSb(111)B \cite{Xu2017,Xu2018}, our data reveals a TSS with a DP located higher in binding energy, which allows to observe the TSS branches below and above the DP. We note that due to the doping level of our \aSn\ films, as well as the limited ARPES resolution, we are not able to observe (gapped) topological surface states in the meV-range near the CB minimum which were recently reported in transport data of HgTe films \cite{mahler2019interplay}.

The two-dimensional character of the TSS is experimentally further confirmed by measuring ARPES at different photon energies [Fig.~\ref{ARPES_standard}(b)]. Clearly, the TSS shows no dispersion with varying photon energy (\kperp\ momentum). At the photon energies corresponding to the bulk $\Gamma$-points in the surface normal direction the intensity of the bulk bands starts to dominate (\kperp\ = 3.3 \invA (2$\times$(2$ \pi / c $)) and 5.0 \invA (3$\times$(2$ \pi / c $))). We note that, similar to the case of the (001)-surface, the momentum distribution of the TSS at given binding energies is isotropic in the \kpara\ plane and does not show any noticeable warping effects [Figs.~\ref{ARPES_standard}(c-e)].

The clear visibility of the DP allows for probing the possible opening of a hybridization gap in TSS upon reducing the \aSn\ film thickness, i.e., to observe a transition from a quasi-3D TI to a quasi-2D TI with a gapped TSS.
Figure \ref{Thickness_ARPES} shows ARPES maps for different \aSn\ film thicknesses. The EDCs taken at $ k_{x} = 0 $ are overlayed on the left side of each map. Despite the relatively thin film thickness of only $ 10 $\,nm, the energy distribution curve (EDC) taken through the DP does not reveal any hybridization gap in the TSS [Fig.~\ref{Thickness_ARPES}(a)].
 
With the thickness reduced to $3$\,nm we observe a slight $p$-doping effect in the electronic structure (energy shift $\Delta E$ $\approx$ 80\,meV). Such a behavior is consistent with the shift of the VB offset of $\sim$100\,meV for thinner \aSn\ films (determined by XPS; not shown here). The  band gap in the projected bulk electronic structure remains unresolved in ARPES as it is still situated above the \Ef. However, a clear reduction of the spectral weight at the DP is observed for the 3-nm-thick film [Fig.~\ref{Thickness_ARPES}(b)]. This is likely a result of the TSS hybridization between the surface- and interface-TSS. 
We can estimate the characteristic decay length of the 2D TSS as $l_{decay} = \hbar v_{TSS}/E_{g} = 1.3$\,nm \cite{linder2009anomalous,zhang2012surface}, where $ E_{g} = 0.7 $\,eV is the gap between the inverted \GEP and \GSM bands in bulk \aSn\ \cite{Rogalev:2016tz}. The surface- and interface-TSS start to significantly hybridize at \aSn\ film thickness of $d_{hybr} = 2\times l_{decay} \approx 2.6$\,nm  which is in good agreement with our data. However, instead of a well-defined hybridization gap, the corresponding EDC taken at the $\Gamma$-point shows a rather broad local minimum at the DP, which could be a combined effect of the background of the projected bulk states and lateral fluctuations in energy positions of the surface- and interface-TSS. The energy difference between the EDC maxima is found to be $\Delta E^{g}_{peak}$ $\approx$ 200\,meV, which is again lower than the theoretical value for a CdTe/\aSn\ quantum well \cite{DeCoster2018}. This discrepancy is most probably due to the more relaxed boundary conditions. 
Interestingly, gapping of the TSS was not resolved in \aSn(001) thin films of similar thickness ($3-4$\,nm) grown on a InSb(001) substrate, yet, for the thicker films ($\approx5$\,nm) a TSS gap of $\sim 200$\,meV was reported \cite{Ohtsubo:2013je}. The latter was attributed to hybridization of the TSS with bulk QW states.

\begin{figure}  [!htb]
	\includegraphics[width=\linewidth]{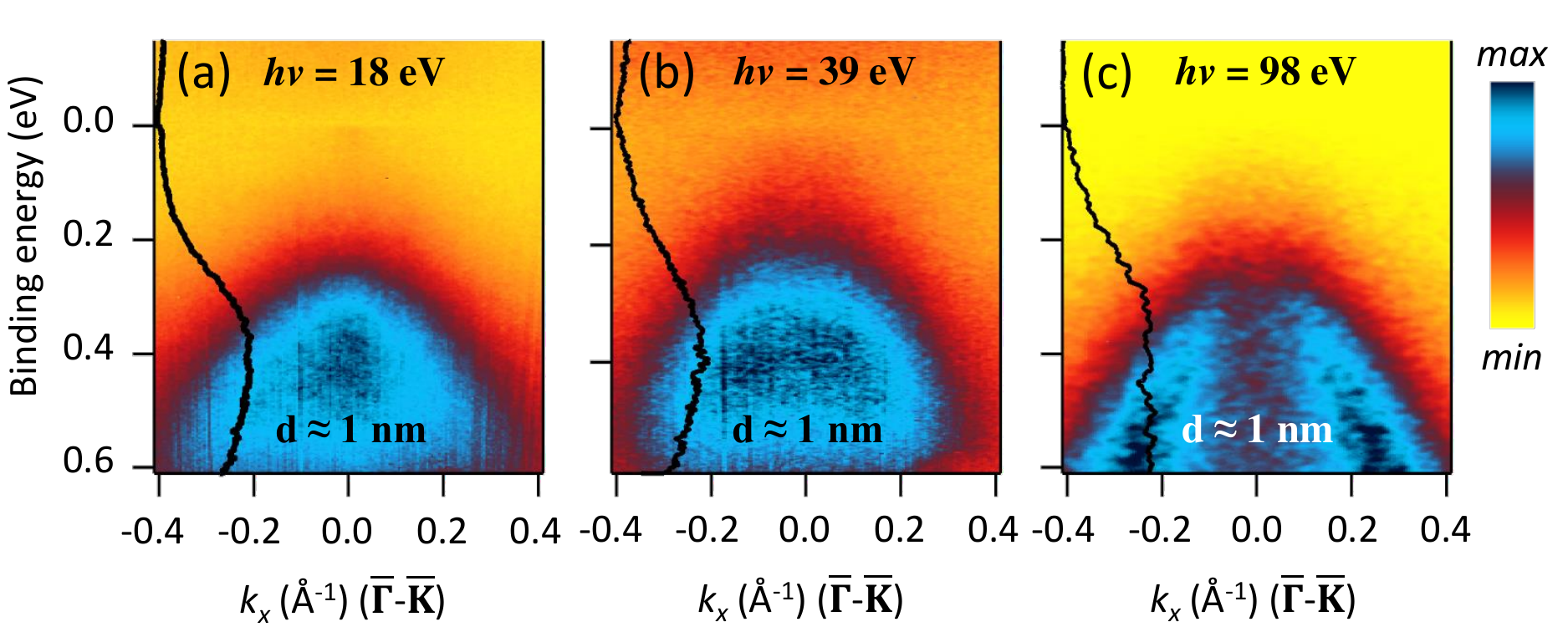}
	\caption{ARPES data measured on a $1$\,nm-thick \aSn\ film at different photon energies $h \nu$ of 18\,eV (left), 39\,eV (center), and 98\,eV (right). EDCs at normal emission are overlayed in each panel. 
	}
	\label{Stanene_ARPES}
\end{figure}

\begin{figure} [ht]
	\includegraphics[width=\linewidth]{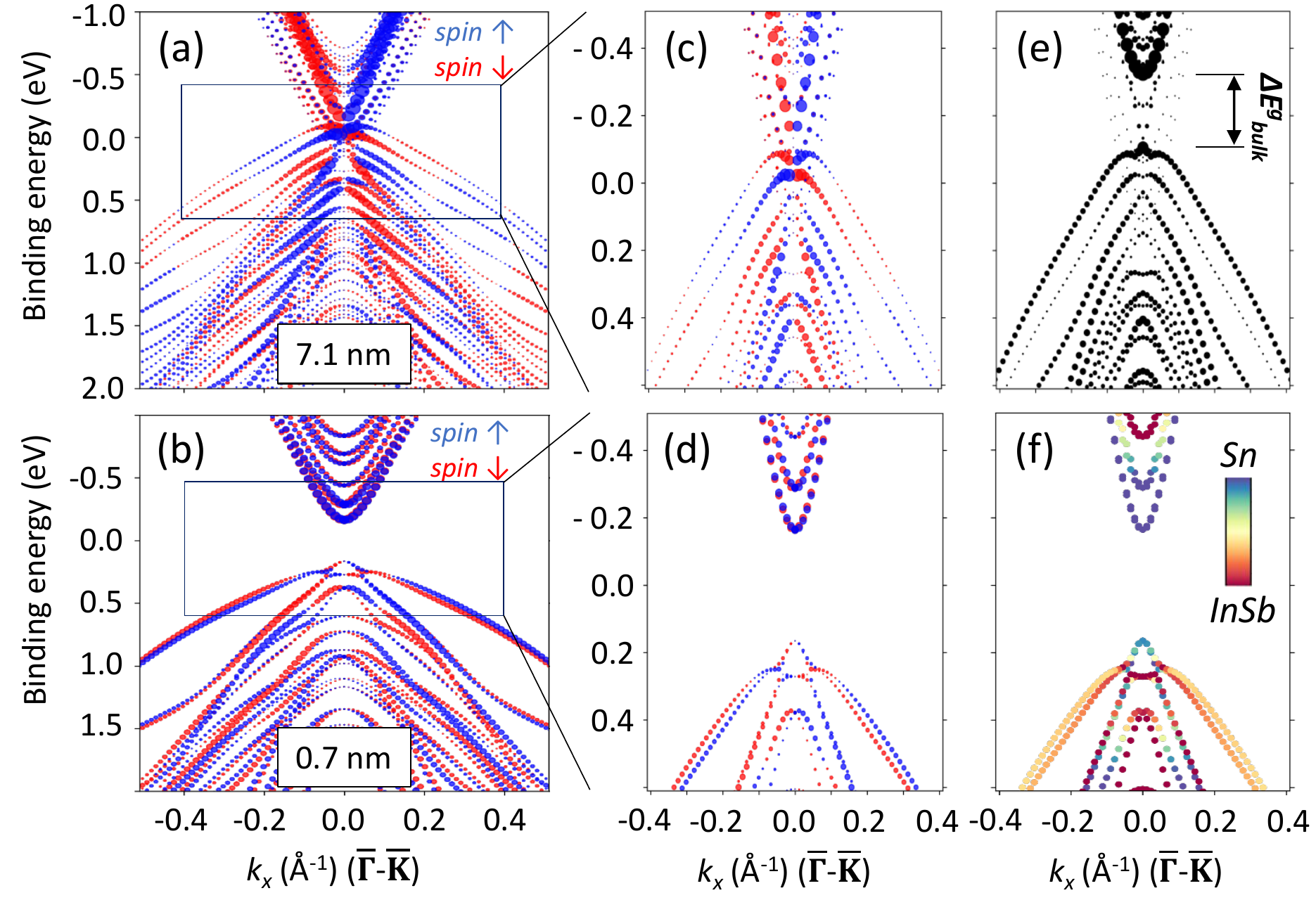}
	\caption{DFT calculations of a $7.1$\,nm-thick (a,c,e) and a $0.7$\,nm-thick (b,d,f) \aSn\ film on InSb(111)A. 
		(a-d) The blue and red colors denote the spin polarization, while the size of the circles is a measure of the \textit{surface} density of states of the \aSn\ film.
		(e) The size of the circles is proportional to the \aSn\ density of states in the middle of the \aSn\ film, i.e., \textit{bulk} states.
		(f) Same as (d), however, here the color indicates the respective contributions from the Sn and InSb states.}
	\label{DFT}
\end{figure}

Reducing the film thickness further, the TSS hybridization becomes stronger. The $1$\,nm-thick \aSn\ film reveals a total gap of $\Delta E^{g}_{peak}$ $\gtrsim$ 400\,meV [see Fig.~\ref{Thickness_ARPES}(c)] since the CB minimum is at or higher than \Ef. In addition, the bulk electronic structure looses dispersive behavior in \kperp\ due to a strong confinement.
This can be seen in Fig.~\ref{Stanene_ARPES} where the \kpara-dispersion shows only a minor variation with photon energy.
As a consequence, we are not able to distinguish experimentally between hybridized TSS states and quasi-2D bulk states.
Note that the measured film data are clearly distinct from those of a clean InSb substrate [Fig.~\ref{Thickness_ARPES}(d)]. The quasi-2D character of the bulk electronic structure of the $1$\,nm-thick film is consistent with the data reported in Ref.\ \cite{Xu2018} for bilayer stanene on InSb(111)B. Nevertheless, our results do not fully agree with the data reported in Refs.\ \cite{Xu2017,Xu2018} for thicker \aSn\ films since the quantum well and TSS hybridization effects are much more pronounced both in our calculations (see Fig.~\ref{DFT}) and in the experimental data.
We note, that at such low \aSn\ film thickness part of the photoelectrons from the substrate can reach the detector and contribute to the observed intensity in the ARPES maps. As a result one can notice a small non-zero photoemission intensity in the gap-region near the $\Gamma$-point in Fig.~\ref{Stanene_ARPES}(b,c) \cite{supplement}.

In order to trace the change of the electronic structure with film thickness we performed DFT calculations. 
Figure \ref{DFT} presents the calculation results obtained for $7.1$\,nm-thick (a,c,e) and $0.75$\,nm-thick (b,d,f) \aSn\ films on InSb(111)A, respectively. The blue and red colors in Figs.~\ref{DFT}(a-d) denote the spin polarization, while the size of the circles is proportional to the density of states in the first two layers for the $0.75$\,nm-thick film, and above the layers for the $7.1$\,nm-thick films. Figures \ref{DFT}(c,d) show the enlarged region around \Ef\, comparable to the experimental data presented in Fig.~\ref{Thickness_ARPES}.

For the $7.1$-nm-thick \aSn\ film, the linearly dispersing spin-polarized surface states can be recognized in the electronic structure [Figs.~\ref{DFT}(a,c)]. Yet, in contrast to the experimental data, the DP is located in the vicinity of \Ef\ [Fig.~\ref{DFT}(c)], which, in turn, is close to the valence band maximum. Apart from this mismatch in the energy position that depends on the alignment of the potentials between the substrate and the \aSn\ film, the theoretical data agrees well with the experimental TSS shown in Fig.~\ref{ARPES_standard}(a). The size of the dots in Fig.~\ref{DFT}(e) is proportional to the density of states in the centre of the \aSn\ film, and thus reveal a quantum confinement gap in the \textit{bulk states} of $\Delta E^{g}_{bulk}$ $\approx$ 200\,meV. The hybridization gap in the TSS [Fig.~\ref{DFT}(c)] has a much smaller value of $\Delta E^{g}_{peak}$ $\approx$ 30\,meV \cite{supplement}.

For the $0.7$\,nm-thick \aSn\ film, a gap of $\sim$330\,meV in the electronic structure is clearly resolved in the calculations [Figs.~\ref{DFT}(b,d)], which agrees well with the experimentally determined gap of at least $\sim$400\,meV [Fig.~\ref{Thickness_ARPES}(c)]. We note that for such thin \aSn\ films, the calculated density of states in the first two layers also contain a contribution from the InSb substrate states. However, as can be seen in Fig.~\ref{DFT}(f), the calculated gap is mainly due to Sn states.

\begin{table}
	[ht] \caption{Experimental TSS hybridization gap and theoretical \textit{s}-like and \textit{p}-like band order near \Ef\ at $\Gamma$-point as a function of \aSn\ thickness in the \aSn/InSb(111)A system.} \label{tab:summary}
	\begin{center}
		\begin{tabular}{c c c}
			\hline
			\hline
			\aSn\ thickness & TSS gap $\Delta E^{g}_{peak}$ & Band order \\
			\hline			
			$\sim$10\,nm & Not resolved & Inverted \\
			$\sim$3\,nm &$\sim$200\,meV & Inverted \\
			$\sim$1\,nm & $\gtrsim$400\,meV & Trivial \\
			\hline
			\hline  	
		\end{tabular}
	\end{center}
\end{table}

We further check the topological character of \aSn\ thin films by tracking the relative energy positions of \textit{s}-like and \textit{p}-like bands near \Ef\ as a function of thickness. The results are listed in Tab.~\ref{tab:summary} together with the values of the TSS hybridization gap estimated from the ARPES data. For the $7.1$\,nm-thick films the band order is inverted, i.e., a \textit{s}-like minimum is below a \textit{p}-like maximum. The band inversion remains also in $2.5$\,nm-thick films where we find that the \textit{s}-like band penetrates into the InSb substrate thus enhancing the effective quantum well width. However, in the $0.7$\,nm-thick \aSn\ film the \textit{s}-like band position is now \textit{above} the \textit{p}-like band, which is a signature of a topologically \textit{trivial} electronic structure. Therefore, the theoretical critical thickness for the transition to a trivial 2D insulator $d_{crit}$ is between $0.7$\,nm and $2.5$\,nm and can be roughly estimated to have a value of $d_{crit} = 1.6$\,nm.
We note that the larger energy difference between \textit{s}-like \GSM and \textit{p}-like \GEP bands in \aSn, as compared to HgTe, leads to a different critical quantum well thickness below which the band inversion is lost. The electron-like band from the \textit{gapped TSS} was calculated to cross the heavy-hole band near the Fermi level in CdTe/\aSn(111) quantum wells at a critical thickness of 2.7\,nm \cite{DeCoster2018} (compare to $ t_{crit} $ = 6.3\,nm for HgTe), defining the transition to a 2D trivial state. 
The topologically trivial band structure was also reported for $ 12 $\,ML-thick ($ \sim 2 $\,nm) \aSn(001) films grown on InSb(001) substrate \cite{Ohtsubo:2013je}.
Moreover, the QSHI phase in free-standing \aSn(001) films was calculated for thickness above $d_{crit}\approx$ 2\,nm \cite{li2019quantum}. In contrast to CdTe-based and free-standing (vacuum-based) quantum wells, the semiconducting InSb substrate allows for less localized interface-TSS at similar \aSn\ film thickness, which, in turn, reduces the TSSs hybridization and quantum well critical thickness.

Finally, we consider the quantum well confinement of the 2D TSSs that exist on the side planes parallel to the sample surface normal of the \aSn\ film. From the uncertainty principle we can approximately estimate the film thickness at which the uncertainty in the quasiparticle momentum is bigger than the typical 2D TSS momentum ($k_{typ} = k_{F} \sim 0.02$\,\invA, see Fig.~\ref{Thickness_ARPES})  $d_{edge} \geq 1/k_{typ} \approx 5$\,nm, which defines the transition of the 2D TSS to the helical 1D edge states. Thus, for the \aSn\ films with thickness between $d_{crit}$ and $\sim$10\,nm, one can expect the presence of a topologically protected 1D helical edge states, while the 2D TSSs remain gapped.

\section{\label{Conclude}Conclusions}
In conclusion, we report on the electronic structure evolution of the TSS in MBE-grown \aSn\ films on InSb(111)A as a function of film thickness. We observe a new (8$ \times $8) surface reconstruction, which was not reported so far. 
As in case of \aSn\ on InSb(001), in $10$-nm-thick \aSn\ films the observed TSS is largely degenerate with the bulk band structure. However, the DP was found to be $\sim$200\,meV below the \Ef. This allows for the observation of a hybridization gap opening in the TSS for thinner \aSn\ films: in $3$\,nm-thick \aSn\ films we determine a gap in the TSS of the order of 200\,meV. The crossover from the 3D to the quasi-2D stanene-like electronic structure in a 1\,nm-thick film is accompanied by a full gap opening ($\Delta E^{g}_{peak}$ $\gtrsim$ 400\,meV) at the $\Gamma$-point in agreement with the calculated few-layer-stanene electronic bandstructure. Our DFT electronic structure calculations of thin \aSn\ films on InSb(111)A show good agreement with the experimental data, as well as provide evidence for the spin-polarized character of the observed TSS. The latter, however, needs to be verified in future experiments. Furthermore, while we have identified the topologically non-trivial character of $\sim$10\,nm- and $\sim$3\,nm-thick \aSn\ films, we find no band inversion in the $\sim$1\,nm-thick \aSn\ film. Therefore, the thickness regime of \aSn\ films where both the gapped TSS and topologically non-trivial bandstructure coexist is between $\sim$2\,nm and $\sim$10\,nm. In addition, this thickness regime corresponds to the appearance of topologically protected 1D helical edge states.

\section{\label{Aknowledge}Acknowledgments}
Valuable   scientific   discussion   with   L. Veyrat and R. St{\"u}hler are acknowledged. G.B. gratefully acknowledges computing time at the JURECA supercomputer of the Jülich Supercomputing Centre (JSC). This work was supported by the Deutsche Forschungsgemeinschaft (DFG) through SPP 1666 Priority Program ”Topological Insulators“, the DFG Collaborative Research Center SFB 1170 ”ToCoTronics” in W\"urzburg and the Würzburg-Dresden Cluster of Excellence on Complexity and Topology in Quantum Matter -- \textit{ct.qmat} (EXC 2147, project-id 39085490). Diamond Light Source (Didcot, UK) is gratefully acknowledged for beamtime under proposals SI10244, SI10289, SI12892, and SI15285. 
LBD acknowledges financial support from EPSRC (UK) and the Science and Technology Facilities Council (UK).

\bibliographystyle{apsrev4-1}

\end{document}



\title{Tailoring the topological surface state \\ in ultrathin \aSn\ (111) films\\ ---Supplementary material---}

\author{V. A. Rogalev}
\affiliation{\mbox{Physikalisches Institut and W\"urzburg-Dresden Cluster of Excellence  ct.qmat, Universit\"at W\"urzburg, 97074 W\"urzburg, Germany}}

\author{F. Reis}
\affiliation{\mbox{Physikalisches Institut and W\"urzburg-Dresden Cluster of Excellence  ct.qmat, Universit\"at W\"urzburg, 97074 W\"urzburg, Germany}}

\author{F. Adler}
\affiliation{\mbox{Physikalisches Institut and W\"urzburg-Dresden Cluster of Excellence  ct.qmat, Universit\"at W\"urzburg, 97074 W\"urzburg, Germany}}

\author{M. Bauernfeind}
\affiliation{\mbox{Physikalisches Institut and W\"urzburg-Dresden Cluster of Excellence  ct.qmat, Universit\"at W\"urzburg, 97074 W\"urzburg, Germany}}

\author{J. Erhardt}
\affiliation{\mbox{Physikalisches Institut and W\"urzburg-Dresden Cluster of Excellence  ct.qmat, Universit\"at W\"urzburg, 97074 W\"urzburg, Germany}}

\author{A. Kowalewski}
\affiliation{\mbox{Physikalisches Institut and W\"urzburg-Dresden Cluster of Excellence  ct.qmat, Universit\"at W\"urzburg, 97074 W\"urzburg, Germany}}

\author{M. R. Scholz}
\affiliation{\mbox{Physikalisches Institut and W\"urzburg-Dresden Cluster of Excellence  ct.qmat, Universit\"at W\"urzburg, 97074 W\"urzburg, Germany}}

\author{L. Dudy}
\affiliation{\mbox{Physikalisches Institut and W\"urzburg-Dresden Cluster of Excellence  ct.qmat, Universit\"at W\"urzburg, 97074 W\"urzburg, Germany}}

\author{L. B. Duffy}
\affiliation{Clarendon Laboratory, Physics Department, Oxford University, OX1~3PU, United Kingdom}

\author{T. Hesjedal}
\affiliation{Clarendon Laboratory, Physics Department, Oxford University, OX1~3PU, United Kingdom}

\author{M. Hoesch}
\affiliation{Diamond Light Source, Didcot, OX11~0DE, United Kingdom}
\affiliation{DESY Photon Science, Notkestraße 85, D-22607 Hamburg}

\author{G. Bihlmayer}
\affiliation{Peter Gr\"unberg Institut, Forschungszentrum J\"ulich and JARA, 52425 J\"ulich, Germany}

\author{J. Sch{\"a}fer}
\affiliation{\mbox{Physikalisches Institut and W\"urzburg-Dresden Cluster of Excellence  ct.qmat, Universit\"at W\"urzburg, 97074 W\"urzburg, Germany}}

\author{R. Claessen}
\affiliation{\mbox{Physikalisches Institut and W\"urzburg-Dresden Cluster of Excellence  ct.qmat, Universit\"at W\"urzburg, 97074 W\"urzburg, Germany}}


\maketitle
\section{Samples characterization}

\begin{figure*} [!hbtp]
	
	\includegraphics[width=0.5\linewidth]{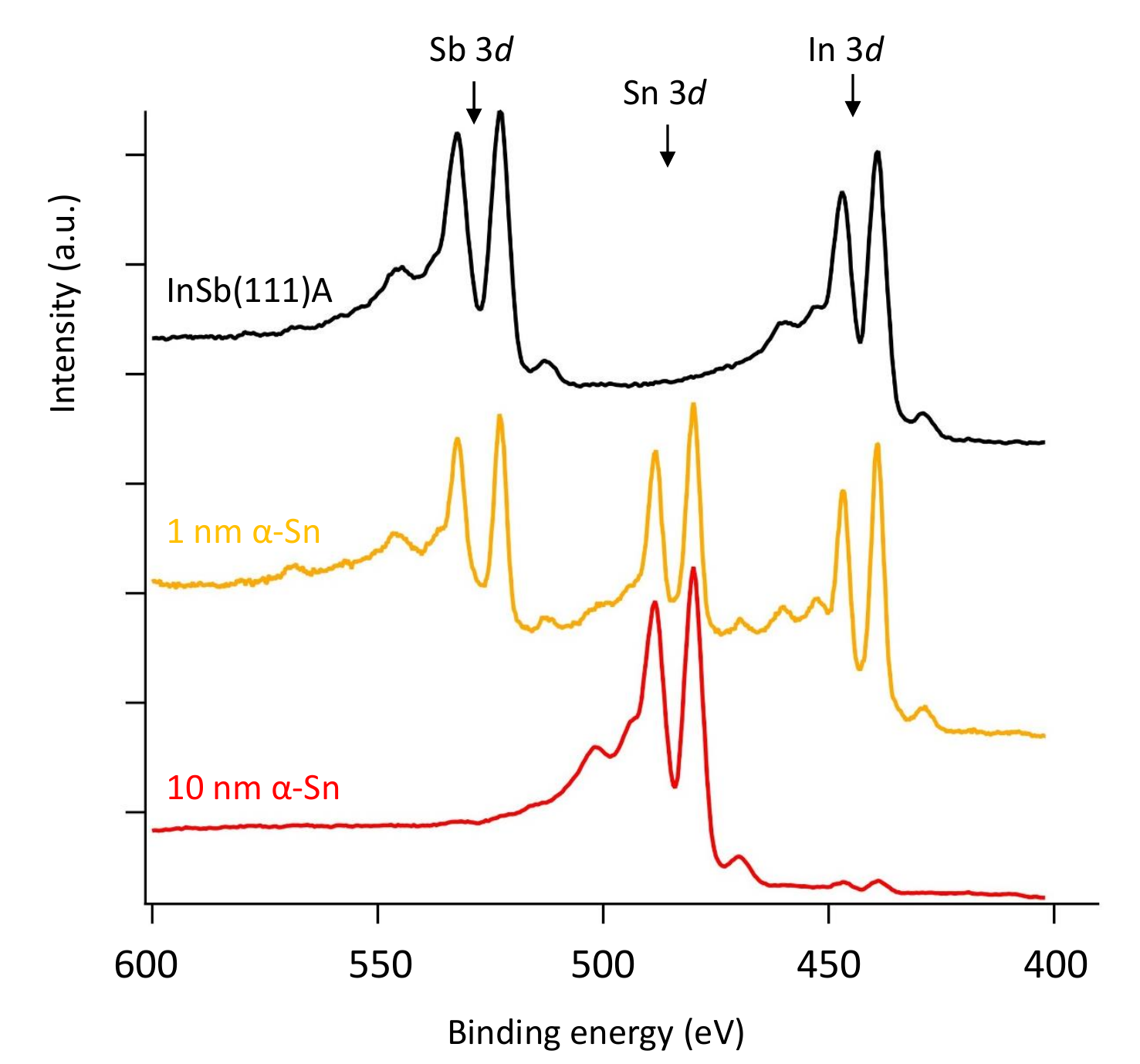}
	
	\caption{Core level spectra measured at $ h\nu $=1486.6 eV (Al K$\alpha$) from the \graySn films with different thickness (red, orange) and from the bare clean InSb(111)A substrate (black).}	
	
	\label{XPS}
\end{figure*}

Fig.~\ref{XPS} shows the core level photoemission spectra from \graySn films with different thickness. 
Spectrum from the $10$\,nm-thick \aSn\ film clearly reveals signal from the In3$ d $ states, which could be a result of In interdiffusion and/or In surface segregation on the substrate.
 
\section{Additional experimental ARPES data measured with different light polarization}

\begin{figure} [!hbt]
	
	\includegraphics[width=0.7\linewidth]{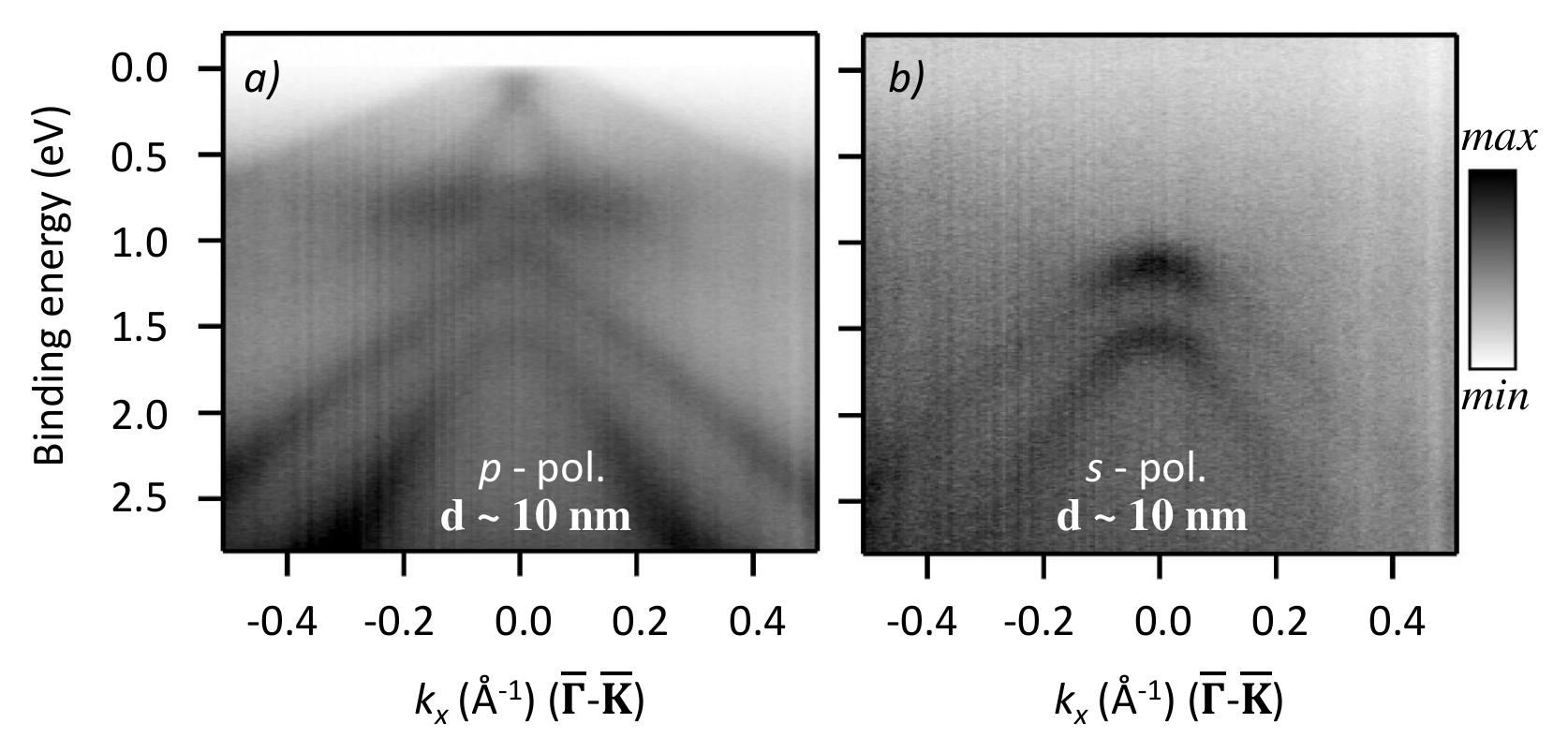}
	
	\caption{a) ARPES maps $I(E,k_{x}$) measured at $ h\nu =$ \unit[18]{eV} ($p$-polarized) along $\overline{\Gamma}\, \overline{\mathrm{K}}$ direction in $10$\,nm-thick \aSn\ film; b) same as a) but measured with $s$-polarized light.}
	
	\label{ARPES_TSS_pol}
	
\end{figure}

Fig.~\ref{ARPES_TSS_pol} presents ARPES maps measured on $10$\,nm-thick \aSn\ film with the $p$ and $s$ linearly polarized light at the photon energy $ h\nu =$ \unit[18]{eV}. The main bulk bands dispersing at higher binding energies are visible in both polarizations, however, the TSSs are visible only in the $p$-polarized light at given experimental geometry.

\begin{figure} [!hbt]
	
	\includegraphics[width=\linewidth]{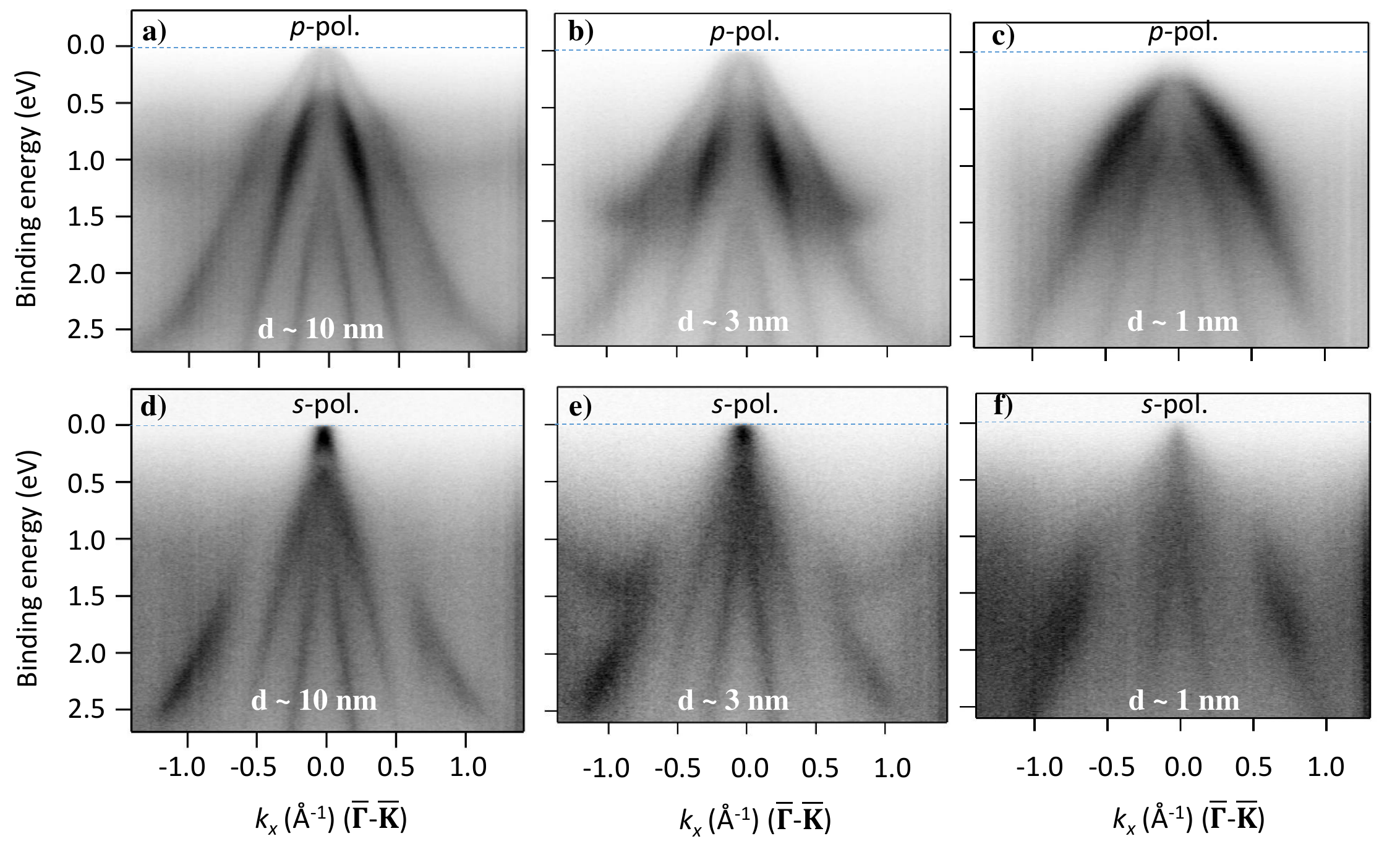}
	
	\caption{ ARPES maps $I(E,k_{x}$) measured on \aSn\ thin films at $ h\nu =$ \unit[98]{eV} using a)-c) $p$-polarized light and d)-f) $s$-polarized light.}
	
	\label{ARPES_3D_pol}
	
\end{figure}

Fig.~\ref{ARPES_3D_pol} presents ARPES maps measured on \aSn\ films with the $p$ and $s$ linearly polarized light at the photon energy $ h\nu =$ \unit[98]{eV} which corresponds to the $\Gamma$-point in the surface perpendicular direction. 
Interestingly, there is a clear intensity redistribution in the valence bands in  $1$\,nm-thick \aSn\ film in comparison to a $10$\,nm- and $3$\,nm-thick films in $p$-polarized light, which is most likely due to the orbital character changes in the band structure. For example, the second valence band in $10$\,nm- and $3$\,nm-thick \aSn\ films has a pronounced intensity maxima in $k_{x} \sim \pm$0.25\,{\invA}(Fig.~\ref{ARPES_3D_pol}a,b), however, in $1$\,nm-thick \aSn\ film these maxima are observed in the first valence band (Fig.~\ref{ARPES_3D_pol}c). 
In addition, in $1$\,nm-thick \aSn\ film there is a clear reduction of intensity at $k_{x} \sim $ 0\,{\invA} in the first and second valence bands.

\begin{figure} [!hbt]
	
	\includegraphics[width=0.7\linewidth]{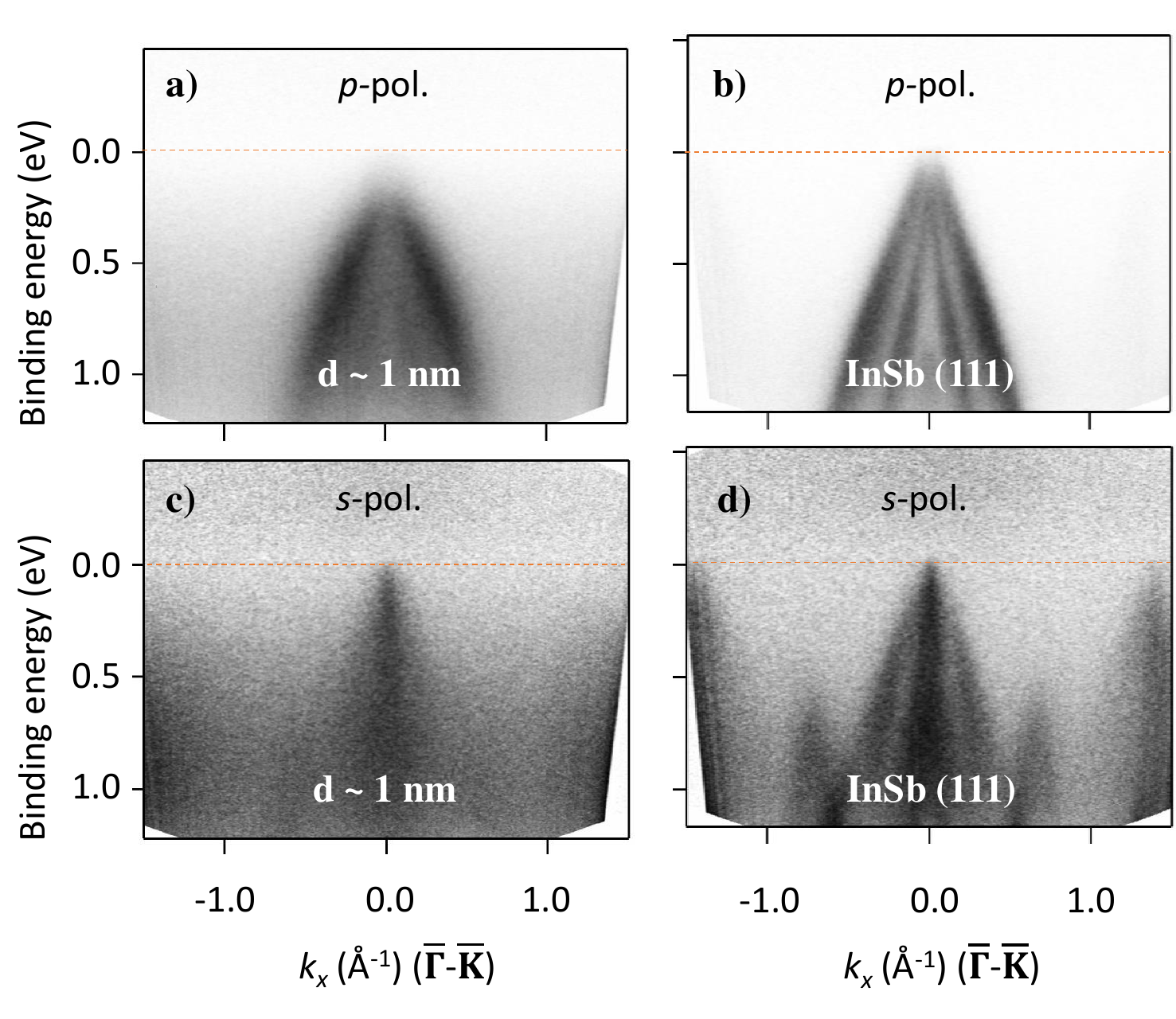}
	
	\caption{ A close-up ARPES maps $I(E,k_{x}$) measured at $ h\nu =$ \unit[98]{eV} on $1$\,nm-thick \aSn\ film (a,c) and clean InSb(111)A substrate (b,d) using $p$- and $s$-polarized light.}
	
	\label{ARPES_2D_pol}
	
\end{figure}

The $1$\,nm-thick \aSn\ film appears to be gapped in the $p$-polarized light (Fig.~\ref{ARPES_3D_pol}c and Fig.4 from the main text), yet, in $s$-polarized light there is a certain photoemission intensity near the Fermi level, which we attribute to the photoelectrons from the InSb(111)A substrate (see Fig.~\ref{ARPES_2D_pol}). In the $s$-polarized light the intensity from the substrate seems to dominate near the Fermi level (hence, no gap), while in $p$-polarized light clearly the \aSn\ film band structure has major contribution. The overlap of the ARPES signals from the substrate and \aSn\ film in $p$-polarized light could explain the intensity drop near $k_{x} \sim $ 0\,{\invA}. Apart from it, the photoemission final-state and matrix-elements effects might play a certain role in the observed ARPES maps. For this reasons the detailed orbital composition analysis of $1$\,nm-thick \aSn\ film is complicated and we leave it for the future studies.

\section{Additional DFT calculation results}

\begin{figure} [!hbt]
	
	\includegraphics[width=0.9\linewidth]{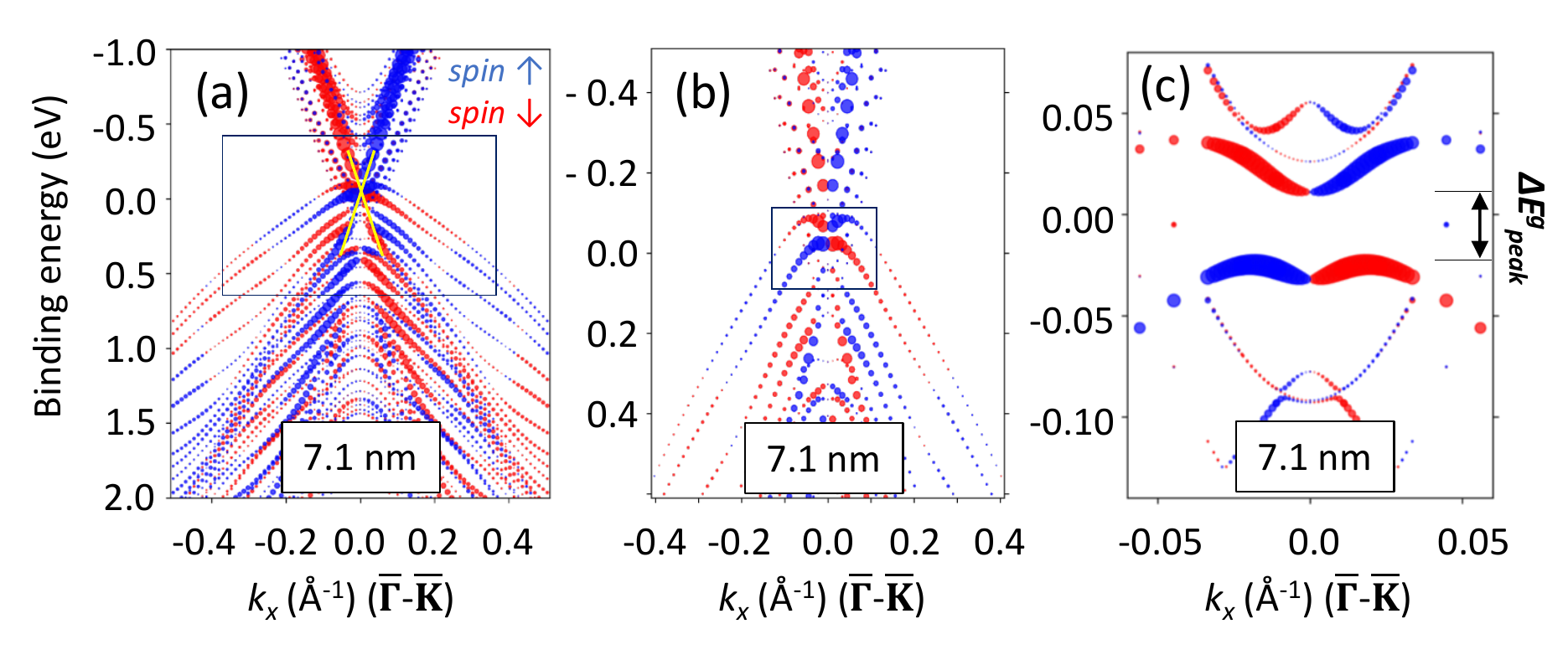}
	
	\caption{ DFT calculations of a $7.1$\,nm-thick \aSn\ film on InSb(111)A at different energy and momentum scale. Yellow lines on panel (a) schematically show TSS.}
	
	\label{DFT_TSS_gap_7nm}
	
\end{figure}

Fig.~\ref{DFT_TSS_gap_7nm} presents DFT calculations of a $7.1$\,nm-thick \aSn\ film on InSb(111)A at different energy and momentum scale in order to resolve hybridization gap $\Delta E^{g}_{peak}$ in TSS. The latter is justified by the spin-polarization swap below and above the gap, which is clearly visible in Fig.~\ref{DFT_TSS_gap_7nm}(c). The hybridization gap $\Delta E^{g}_{peak}$ is found to be $\approx$ 30\,meV.